\documentclass[numberedappendix,useAMS,usenatbib,letterpaper]{mn2e}
\usepackage[pass]{geometry}
\usepackage{amssymb,amsmath}
\usepackage{verbatim}
\usepackage{color,hyperref}
\usepackage{needspace}
\usepackage{enumitem}
\usepackage{tikz}
\usepackage{graphicx}
\usetikzlibrary{positioning}
\definecolor{linkcolor}{rgb}{0,0,0.25}
\hypersetup{
  colorlinks=true,        
  linkcolor=linkcolor,    
  citecolor=linkcolor,    
  filecolor=linkcolor,    
  urlcolor=linkcolor      
}
\newcounter{address}
\title[The $m=4$ bar]
{The 4:1 Outer Lindblad Resonance of a long slow bar as a potential explanation for the Hercules stream}

\author[J. A. S. Hunt \& J. Bovy]
{\parbox{\textwidth}{Jason A.~S.~Hunt$^1$ \& Jo~Bovy$^{1,2,3}$}\vspace{0.5cm}
\\
$^{1}$ Dunlap Institute for Astronomy and Astrophysics, University of Toronto, 50 St. George Street, Toronto, Ontario, M5S 3H4, Canada\\
$^{2}$ Department of Astronomy and Astrophysics, University of Toronto, 50 St. George Street, Toronto, ON, M5S 3H4, Canada \\
$^{3}$ Alfred P. Sloan Fellow\\
}

\pagerange{\pageref{firstpage}--\pageref{lastpage}}
\pubyear{2017}

\begin{document}

\maketitle

\label{firstpage}

\begin{abstract}
There are multiple groups of comoving stars in the Solar neighbourhood, which can potentially be explained as the signatures of one of the fundamental resonances of non-axisymmetric structure such as the Galactic bar or spiral arms. One such stream, Hercules, has been proposed to result from the outer Lindblad resonance (OLR) of a short fast rotating bar as shown analytically, or the corotation resonance (CR) of a longer slower rotating bar as observed in an $N$-body model. We show that by including an $m=4$ Fourier component in an analytical long bar model, with an amplitude that is typical for bars in $N$-body simulations, we can reproduce a Hercules like feature in the kinematics of the Solar neighbourhood. We then describe the expected symmetry in the velocity distribution arising from such a model, which we will soon be able to test with $Gaia$.
\end{abstract}


\section{Introduction}\label{intro}
The kinematics of the Solar neighbourhood shows rich substructure in the form of streams or moving groups. One such stream, Hercules, consists of stars with $U\sim-30$ km s$^{-1}$, and $V\sim-50$ km s$^{-1}$ with respect to the Sun's velocity, where $U$ is velocity in the direction of the Galactic centre, and $V$ is velocity in the direction of rotation. It has been proposed that Hercules is a result of the Outer Lindblad Resonance (OLR) of the Galactic bar \citep{D00}. This arrises naturally from resonant interaction between the disc and a short, fast rotating bar, as further explored in e.g. \citet{Aetal14-2,Monari+16}. 

However, if the bar is longer, as favoured by more recent measurements of its extent \citep[e.g.][]{WGP15}, then the bar must rotate slower, because it may not extend past corotation \citep[e.g.][]{Contopoulos80}. For such a bar, the OLR is located at around 10.5 kpc, making it unable to account for the Hercules stream \citep[although note that a possible OLR feature has been observed around 10-11 kpc in][]{Liu+12}. 

\cite{P-VPWG17} present an alternate explanation for the Hercules stream arising from a bar with a half length of 5 kpc. In this model, stars orbiting the bar's Lagrange points $\mathcal{L}_4$ and $\mathcal{L}_5$ move outward from corotation which occurs at $R=6$ kpc to reach the Solar neighbourhood. However, explorations of the distribution function in the outer disc for a long slow bar model, either show no Hercules like feature \citep[e.g.][]{2017MNRAS.465.1443M}, or one which is substantially weaker \citep{MFFB17,Binney2018} than is seen in either the model from \cite{P-VPWG17}, or the Solar neighbourhood. These models typically use a simple quadrupole bar potential \citep[e.g. from][]{D00}, which is a simple approximation for the complex structure in the inner Galaxy. 

While the exact morphology of the bar is not yet fully constrained, various studies favour a more complicated potential than is given by the Dehnen bar. For example, the Milky Way appears to host an X-shaped bulge \citep[e.g.][]{NL16} such as is often seen in external galaxies \cite[e.g][]{LDP00} and can be easily created in $N$-body simulations \citep[e.g.][]{CS81,Athanassoula05,AVSD17}. Fourier decomposition of the density distribution in simulated \citep[e.g.][]{AM02} and observed \citep[e.g.][]{QFG94,BLSBK06,D-GSL16} barred disc galaxies show evidence for the existence of higher order modes than the $m=2$ Dehnen bar. Thus, it is possible that the inability of the long-slow bar scenario to make a convincing analytical prediction of the Hercules stream arises from the simplicity of the assumed bar potential.

In this work, we show that a long slow bar such as is shown in \cite{P-VPWG17} is capable of producing a Hercules-like feature in the Solar neighbourhood $UV$ plane, when including an $m=4$ Fourier component in the model for the bar potential. 

In Section \ref{simulation} we present seven $N$-body galaxy models which show varying bar morphologies, and compute the $m=4$ Fourier mode of the bars. In Section \ref{modelling} we present models of the long bar with and without a $m=4$ component and discuss the impact on the resulting velocity distribution, both in the Solar neighbourhood, and further across the Galactic disc. In Section \ref{4sym} we discuss the expectation of symmetry in the velocity distribution across the Galactic disc if the bar has an $m=4$ component. In Section \ref{summary} we summarize our results.

\section{Simulations}\label{simulation}
It is known from observations of external galaxies that within barred spiral disc galaxies the size and shape of the central bar can vary greatly. In this Section we present seven $N$-body models which display different morphological properties and perform a Fourier transform on their central bars to recover the amplitude of the various modes of the radial force.

The set of galaxies were generated with the $N$-body/SPH code \sc{gcd+ }\rm \citep[e.g.][]{KG03}. Most have been examined in other works, in which you will find the details of the numerical simulation code and the setup procedure. Table \ref{Tpars} shows their basic parameters where $M_{200}$ ($M_{\sun}$) is the total halo mass, $M_{\mathrm{d}}$ ($M_{\sun}$) is the mass of the stellar disc, $c$ is the concentration parameter, $R_{\mathrm{d}}$ (kpc) is the scale length of the disc and $z_{\mathrm{d}}$ (kpc) is the scale height of the disc. For models including a thick disc, the parameter for the thick disc is in brackets.

\begin{table}
\centering
\caption{Parameters of the simulations displayed in Fig. \ref{morphology}. For models with a thin and thick disc, the value outside the brackets is the parameter for the thin disc, and the value in the brackets is the parameter for the thick disc.}
\label{Tpars}
\renewcommand{\footnoterule}{}
\begin{tabular}{@{}cccccccccccc@{}}
\hline
Model &  $M_{200}$  & $M_{\mathrm{d}}$  & $c$ & $R_{\mathrm{d}}$  & $z_{\mathrm{d}}$  \\
 & ($10^{12}M_{\sun}$) & ($10^{10}M_{\sun}$) &  & (kpc) & (kpc) \\ \hline
A & 2.5 & 4.0 & 10.0  & 2.5 & 0.35 \\ \hline
B & 2.0 & 5.0 & 9.0 & 3.0 & 0.3  \\ \hline
C & 1.5 & 4.5(1.5) & 12.0  & 4.0(2.5) & 0.35(1.0)   \\ \hline
D & 1.75 & 5.0 & 9.0  & 3.0 & 0.3   \\ \hline
E & 2.5 & 4.0(1.0) & 10.0  & 2.5(2.0) & 0.3(1.0)   \\ \hline
F & 1.0 & 5.5(0.5) & 10.0 & 3.0(2.5) & 0.35(1.0)  \\ \hline
G & 2.0 & 5.0 & 10.0  & 2.0 & 0.3   \\ \hline
\end{tabular}
\end{table}

\begin{figure}
\centering
\includegraphics[width=\hsize]{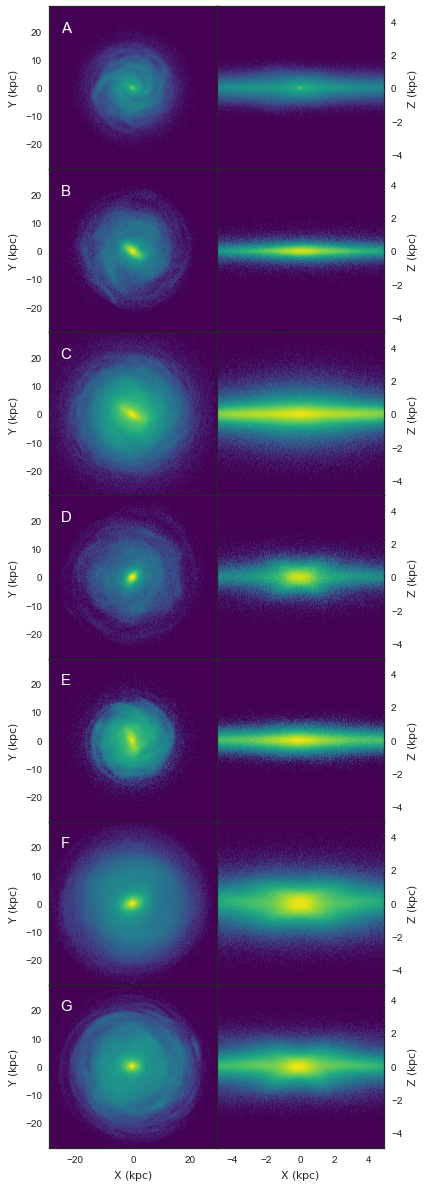}
\caption{Face-on (left) and edge-on (right) view of Models A-G (top to bottom).}
\label{morphology}
\end{figure}

Figure \ref{morphology} shows the face on (left) and edge on (right) morphology of model galaxies A to G (top to bottom). Model A  \citep[which was previously analysed in][]{KHGPC14,HKGMPC15} contains a short flat bar, with strong spiral structure and is comprised of a single disc. Model B \citep[previously analysed as Target II in][]{HKM13} contains a long flat bar, with light spiral structure and is comprised of a single disc. Model C \citep[previously analysed in][]{KGGCHB17} contains a long flat bar, with light spiral structure and is comprised of a thin and thick disc. Model D \citep[previously analysed as Target IV in][]{HKM13} contains a short bar, little spiral structure and is comprised of a single disc. Model E contains a long flat bar and is composed of a thin and thick disc. Model F contains a short bar and is comprised of a thin and thick disc. Model G contains a short bar, light spiral structure and is comprised of a single disc. Models D, F and G have a pronounced X shape when viewed edge on, whereas A, C, D and E have little vertical structure.

\begin{figure}
\centering
\includegraphics[width=\hsize]{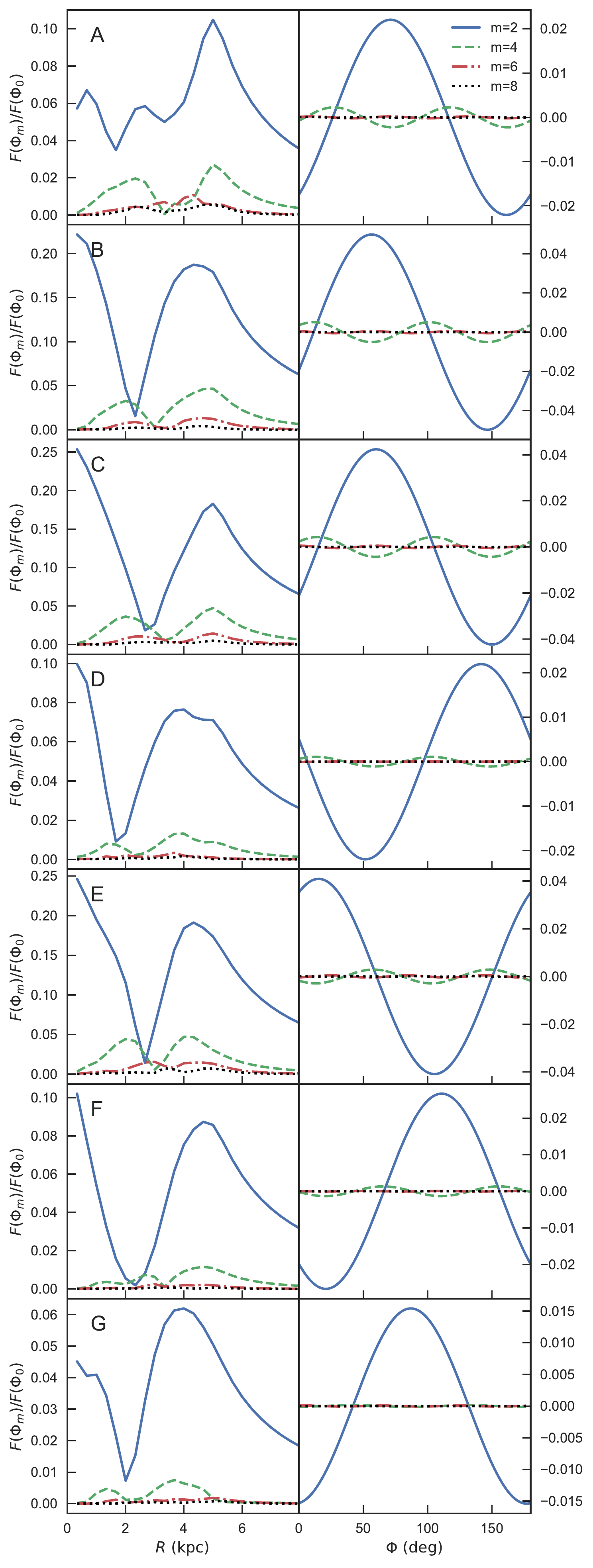}
\caption{Fourier decomposition of the ratio of the radial force $F(\Phi)_m/F(\Phi)_0$ for Models A-G (top to bottom) by galactic radius (left) and by angle at a distance of 8 kpc (right), for the $m=2$ (blue solid), $m=4$ (green dashed), $m=6$ (red dot-dashed) and $m=8$ (black dotted) mode compared to the axisymmetric background potential.}
\label{fourier2}
\end{figure}

Figure \ref{fourier2} shows the Fourier decomposition of the radial force of the bar area, using the simple cut of $R_{\mathrm{G}}\leq5$ kpc, by distance from the galactic centre (left) and by angle at 8 kpc (right) of model galaxies A to G (top to bottom) for $m=2$ (blue solid), $m=4$ (green dashed), $m=6$ (red dash-dot) and $m=8$ (black dotted). The radial force at the Solar radius calculated from the bar region is dominated by the $m=2$ mode, with the force ranging from around 1.5\% - 4\% of the total force coming from the $m=2$ component of the bar, compared with the axisymmetric background at $R_0$. All Models except G have a visible $m=4$ component, albeit substantially weaker than the $m=2$ mode. The force contribution from the $m=4$ component ranges from 0.1\% - 0.4\% of the total force in Models A-F. Model G does contains a magnitude smaller $m=4$ force component with 0.001\% of the total radial force. In general, the amplitude of the $m=4$ component of the radial force is around $5-10$\% of the amplitude of the $m=2$ component. It is also worth noting that each simulation examined here contains a negative $m=4$ component, e.g. the maximum of the $m=2$ mode aligns with a minimum of the $m=4$ mode. We are not suggesting that all $m=4$ components are negative, it is merely the case for this set of models.

Figures \ref{morphology} and \ref{fourier2} do not indicate a strong link between a bar containing significant vertical structure such as a the X shape, and the density or radial force having a large $m=4$ component. However, there is a slight correlation in the strength of the $m=4$ component. For example, Models A, B, C and E, all contain flat bars, and while they are dominated solely by the $m=2$ bar, they contain $m=4$ components of similar amplitudes. Models D, F and G contain X shaped structure, and while they also contain $m=4$ components, in Models D and F the amplitude is slightly less than in the flat bars, and Model G has very little or no $m=4$ component. The correlation here is slight, and it would require further study to determine if such features are more likely to be found together, separately, or entirely uncorrelated. Regardless, we take the presence of $m=4$ components in the flat bar models as sufficient justification to explore the effect of the $m=4$ component in a two dimensional model in Section \ref{modelling}. The models also contain small $m=6$ and $m=8$ components, but at a much lower amplitude.

\section{Modelling the higher order Fourier components of the bar}\label{modelling}
To make predictions of the velocity distribution in the Solar neighbourhood, resulting from resonant interaction with the Galactic bar, we use \texttt{galpy}\footnote{Available at \url{https://github.com/jobovy/galpy}~.} \citep{B15} to simulate the stellar orbits. We choose to investigate the effects of including an $m=4$ component in the bar potential because it is the most significant of the higher order modes, and because for some long-slow bar models the 4:1 OLR occurs around the location of the Sun \citep[e.g. $\sim8$ kpc in][]{LGSPW16}

As mentioned earlier, because the simulations in Section \ref{simulation} show the $m=4$ component can occur in a flat bar, we ignore the vertical motions and only simulate the two-dimensional dynamics in the Galactic plane. As shown in \cite{Bovy10} and \cite{Hunt+18} we use a Dehnen distribution function \citep{Dehnen99} to model the stellar disc before bar formation, and represent the distribution of stellar orbits. This distribution function is a function of energy $E$ and angular momentum $L$
\begin{equation}
f_{\text{dehnen}}(E,L)\ \propto \frac{\Sigma(R_e)}{\sigma^2_{\text{R}}(R_e)}\exp\biggl[\frac{\Omega(R_e)[L-L_c(E)]}{\sigma^2_{\text{R}}(R_e)}\biggr],
\end{equation}
where $R_e$, $L_c$ and $\Omega(R_e)$ are the radius, angular momentum and angular frequency, respectively, of a circular orbit with energy $E$. The gravitational potential is assumed to be a simple power-law, such that the circular velocity is given by
\begin{equation}
  v_c(R)=v_0(R/R_0)^{\beta}\,,
\end{equation}
where $v_0$ is the circular velocity at the solar circle at radius $R_0$.

To model the bar we generalise the simple quadrupole bar potential from \citet{D00} to a general $\cos(m\phi)$ potential such that
\begin{equation}
\begin{split}
&\Phi_{\mathrm{b}}(R,\phi)=A_{\text{b}}(t)\cos(m(\phi_{\mathrm{b}}-\Omega_{\text{b}}t))\\ 
& \quad \quad \times
\left\{ \begin{array}{ll} -(R/R_0)^p, & \mathrm{for}\ R \geq R_{\text{b}},\\ ([R_{\text{b}}/R]^p-2)\times(R_{\mathrm{b}}/R_0)^p, & \mathrm{for}\ R \leq R_{\text{b}}, \end{array}
\right.
\end{split}
\end{equation}
where $R_{\text{b}}$ is the bar radius, set to $80\%$ of the corotation radius, $m$ is the integer multiple of the $\cos$ term, $\phi_{\mathrm{b}}$ is the angle of the bar with respect to the Sun--Galactic-center line and $p$ is the power-law index. To reproduce the Dehnen bar, $m=2$ and $p=-3$. 
The bar is grown smoothly following the prescription
\begin{eqnarray}
A_{\text{b}}(t)=
\left\{\begin{array}{ll} 0,\ \frac{t}{T_{\text{b}}}<t_{\text{1}} \\ A_f\biggl[\frac{3}{16}\xi^5-\frac{5}{8}\xi^3+\frac{15}{16}\xi+\frac{1}{2}\biggr], t_{\text{1}}\leq\frac{t}{T_{\text{b}}}\leq t_{\text{1}}+t_{\text{2}}, \\ A_f,\ \frac{t}{T_{\text{b}}} > t_{\text{1}} + t_{\text{2}}.  \end{array}
\right.\,
\end{eqnarray}
where $t_1$ is the start of bar growth, set to half the integration time, and $t_2$ is the duration of the bar growth. $T_{\text{b}}=2\pi/\Omega_{\text{b}}$ is the period of the bar,
\begin{equation}
\xi=2\frac{t/T_{\text{b}}-t_{\text{1}}}{t_{\text{2}}}-1,
\end{equation}
and
\begin{equation}
A_f=\alpha_{m}\frac{v_0^2}{3},
\end{equation}
where $\alpha_{m}$ is the dimensionless ratio of forces owing to the $\cos(m\phi)$ component of the bar potential and the axisymmetric background potential, $\Phi_0$, at Galactocentric radius $R_0$ along the bar's major axis, corresponding to the amplitude in Figure \ref{fourier2}. Another common method of measuring bar strength is $Q_{\mathrm{r}}$, which is related to $\alpha$ such that $\alpha=Q_{\mathrm{r}}(R_0)$, where 
\begin{equation}
Q_{\mathrm{r}}(r)=\frac{\partial\Phi_{\mathrm{b}}/\partial r}{\partial\Phi_0/\partial r}.
\end{equation}

\begin{figure}
\centering
\includegraphics[width=\hsize]{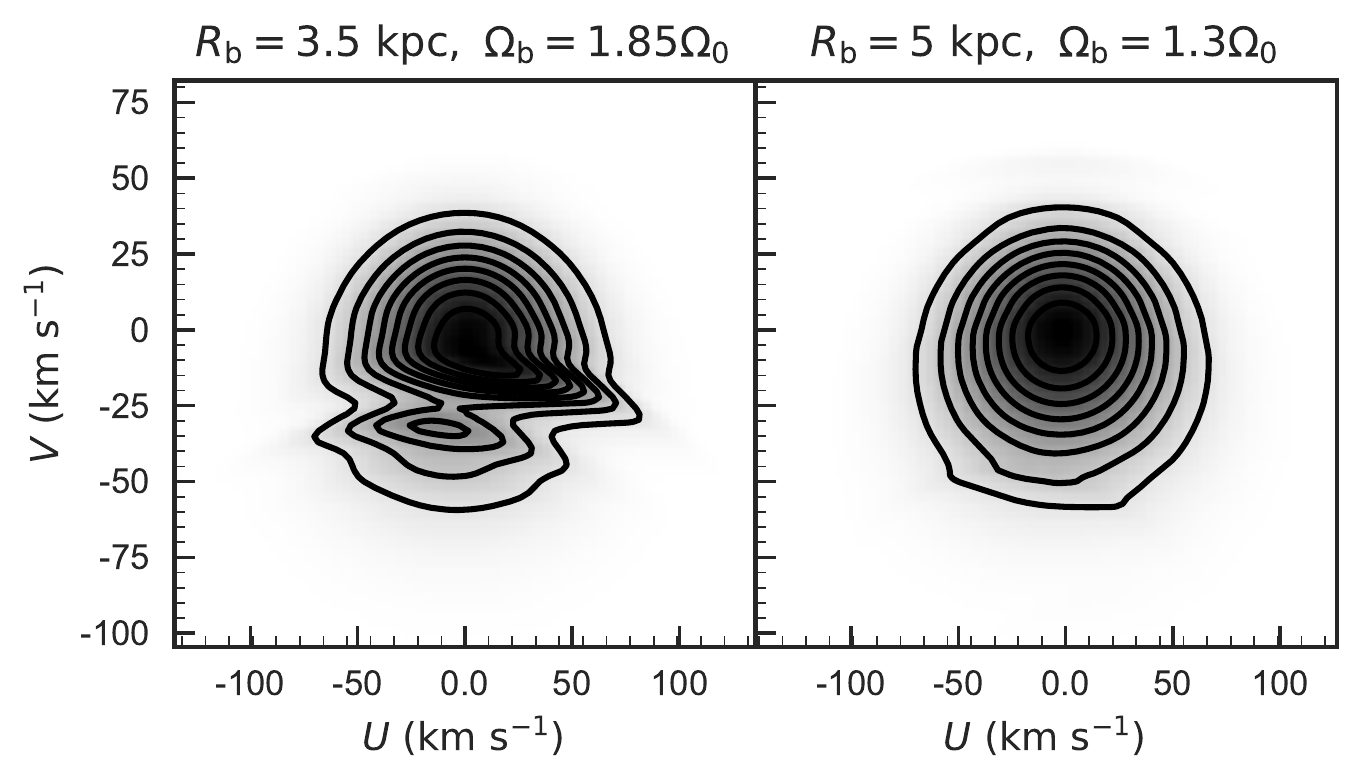}
\caption{$UV$ plane in the Solar neighbourhood for an example of a short fast bar (left) and a long slow bar (right).}
\label{simple}
\end{figure}
\begin{figure}
\centering
\includegraphics[width=\hsize]{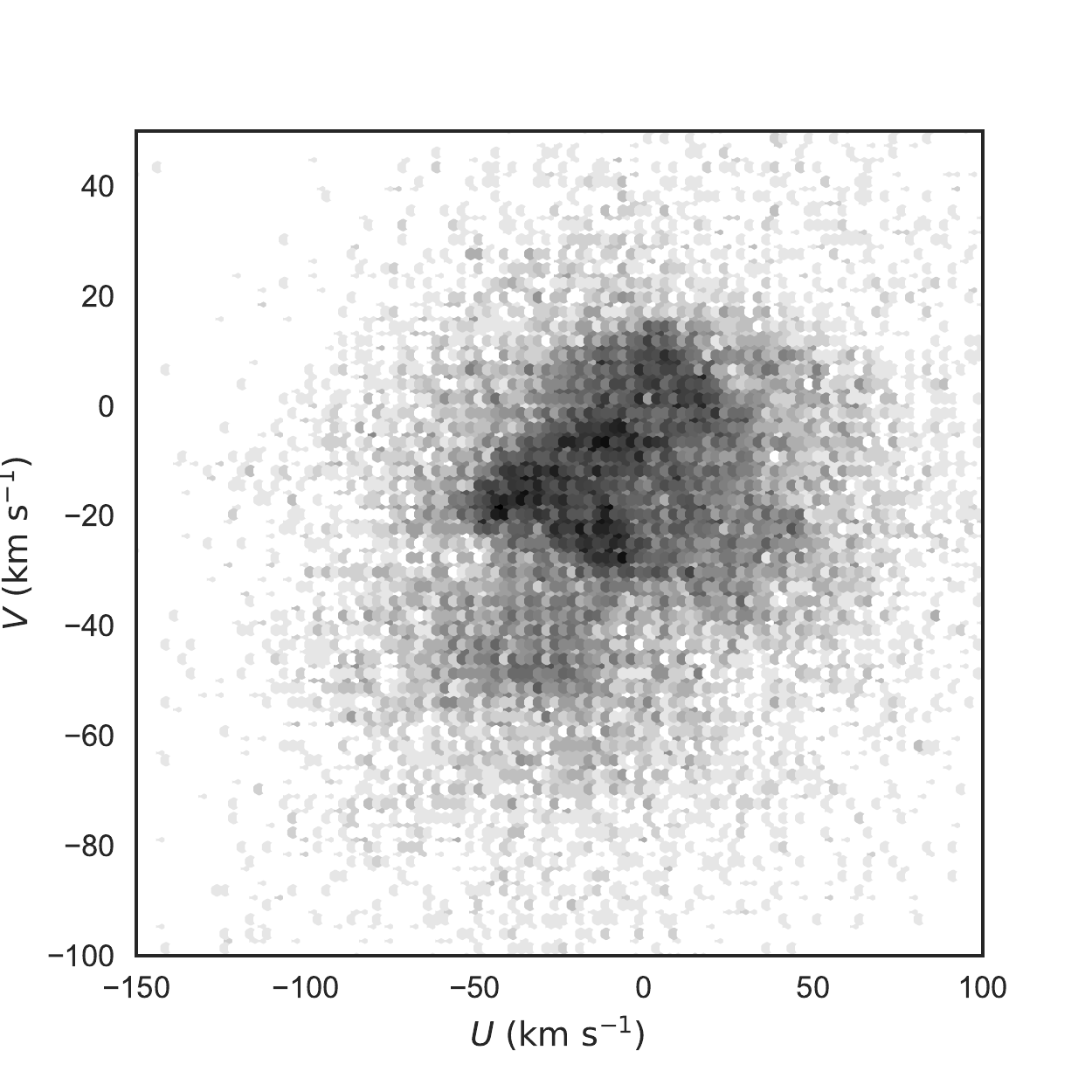}
\caption{$UV$ plane in the Solar neighbourhood for the combined TGAS-RAVE data satisfying $\sigma_{\pi}/\pi\leq0.1$ and $1/\pi<0.2$ kpc.}
\label{tgasrave}
\end{figure}

\begin{figure*}
\centering
\includegraphics[width=\hsize]{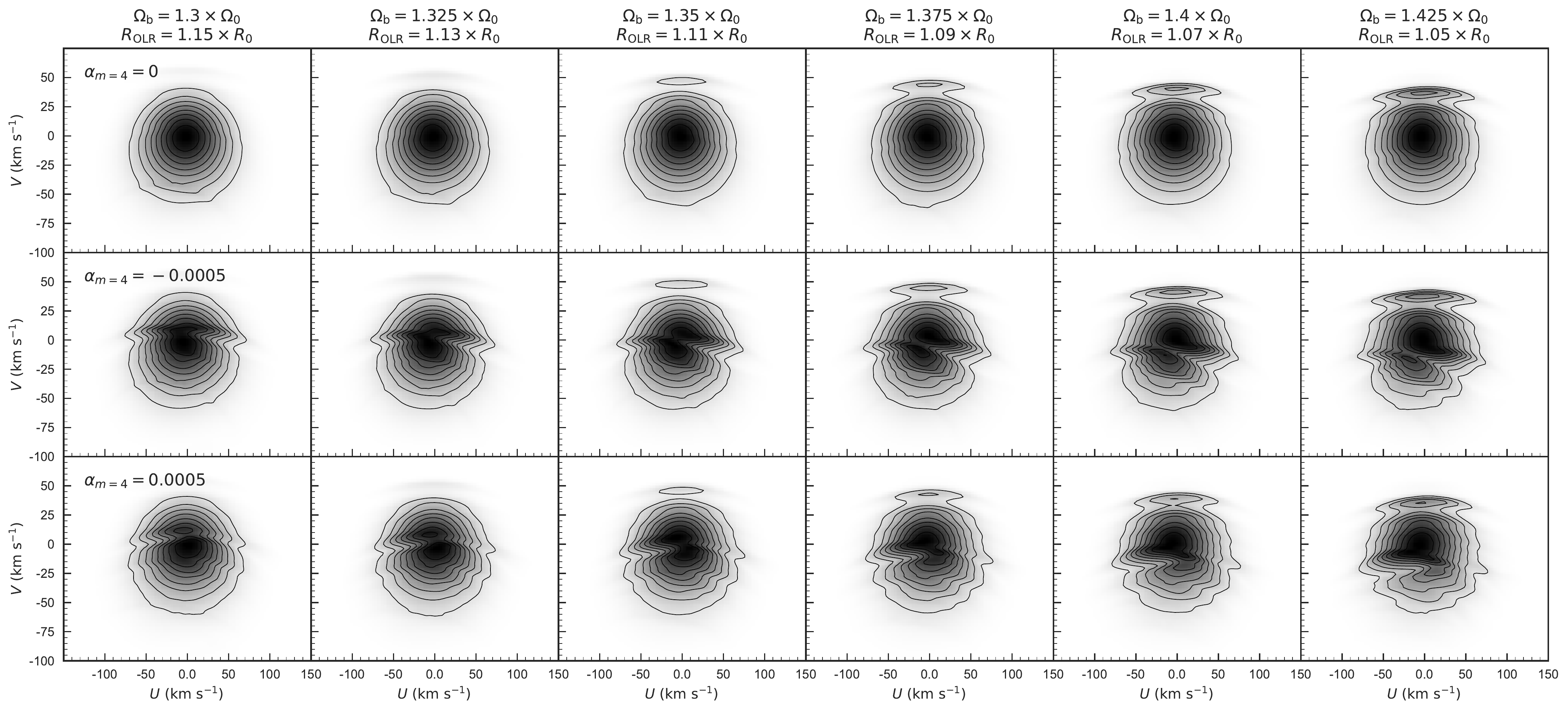}
\caption{$UV$ plane in the Solar neighbourhood for a 5 kpc bar with pattern speed $\Omega_{\mathrm{b}}=1.3\times\Omega_0$ to $\Omega_{\mathrm{b}}=1.425\times\Omega_0$ from left to right for a purely $m=2$ bar (upper row), a bar with a negative $m=4$ component added (centre row) and a bar with a positive $m=4$ component added (lower row).}
\label{m4slow}
\end{figure*}

Figure \ref{simple} shows the $UV$ plane in the Solar neighbourhood for the simple bar model for a short fast bar (left) and a long slow bar (right). For our fast bar model, we set $\alpha_{m=2}=0.01$, $R_0=8.0$ kpc, and $v_0=220$ km s$^{-1}$, the bar has an angle of $25^{\circ}$ with respect to the Sun--Galactic-center line, a pattern speed of $\Omega_{\text{b}}=1.85\times\Omega_0$  and a half length of 3.5 kpc. For the slow bar model we set $\alpha_{m=2}=0.01$, $R_0=8.0$ kpc and $v_0=220$ km s$^{-1}$, the bar has an angle of $25^{\circ}$ with respect to the Sun--Galactic-center line, a pattern speed of $\Omega_{\text{b}}=1.3\times\Omega_0$  and a half length of 5 kpc. For these simple bar models, the fast bar model naturally recreates a strong feature in the $UV$ plane in the region of Hercules, around $U=-30$ km s$^{-1}$ with respect to the LSR, or $U=-40$ km s$^{-1}$ with respect to the Sun, while the slow bar model does not. Note that these models are but a single example of the parameters for a short fast, and long slow bar, and different choices will result in a different $UV$ plane. However, the trend is consistent for short fast rotating bars which have been examined in numerous previous works \citep[e.g.][]{D00,Aetal14-2,Monari+16}.

For comparison, we cross match the Tycho-$Gaia$ Astrometric Solution \citep[TGAS;][]{Michalik+15} catalogue from the European Space Agency's $Gaia$ mission \citep{GaiaMission} with data from the Radial Velocity Experiment \citep[RAVE;][]{Sea06} to attain six-dimensional phase space information for over 200,000 stars in the Solar neighbourhood. We perform a simple cut requiring fractional parallax error $\sigma_{\pi}/\pi\leq0.1$, and stellar distances of $1/\pi\leq0.2$ kpc, resulting in a sample of 26,792 stars. Fig. \ref{tgasrave} shows the $UV$ plane in the Solar neighbourhood for this sample, without any correction for Solar motion. The Hercules stream is clearly visible in the lower left of the figure, around $(U,V)=(-40,-50)$ km s$^{-1}$. The Hyades, Pleiades, Coma Berenices and Sirius moving groups are also visible within the main mode.

Note that neither of the simple models presented in Fig. \ref{simple} reproduce the complex structure in the main peak of the density in the $UV$ plane (e.g. the other moving groups) shown in Fig. \ref{tgasrave}. This is unsurprising considering it is thought to be heavily influenced by interaction with other non-axisymmetric structure such as the spiral arms \citep[e.g.][]{Quillen03,QM05,Sellwood2010,MVB17}.

To include an $m=4$ component of the bar, we add a hexadecapole bar potential. We grow this second potential along with the main bar, assuming the same bar length and rotation, and a radial drop-off of $p=-5$. We compare the $UV$ plane in the Solar neighbourhood for a pure $m=2$ bar against bars with a positive and negative $m=4$ component, although only a negative $m=4$ component is observed in the simulations, e.g. the maximum of the $m=2$ mode aligns with a minimum of the $m=4$ mode.

\begin{figure*}
\centering
\includegraphics[width=\hsize]{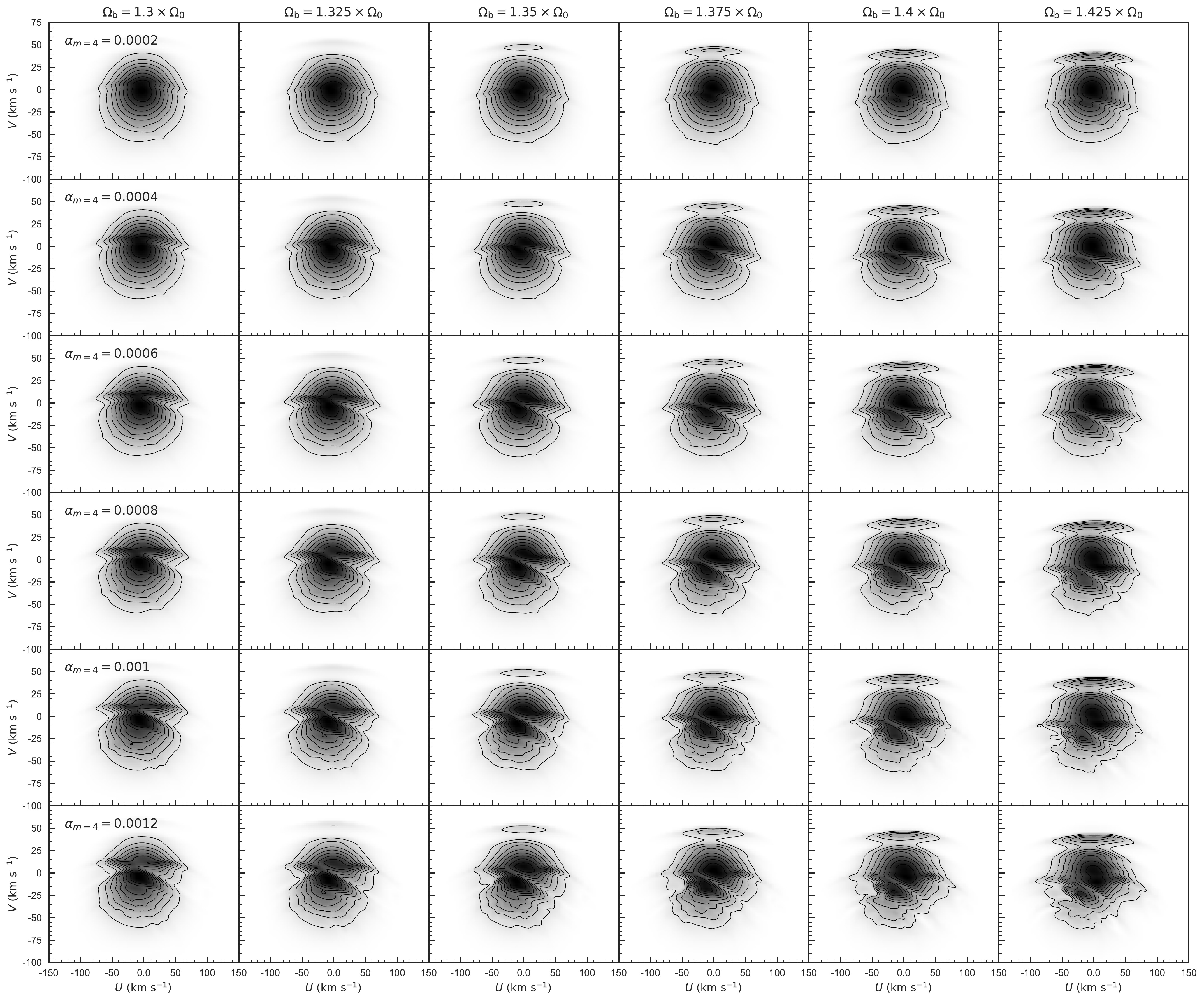}
\caption{$UV$ plane in the Solar neighbourhood for a 5 kpc bar with pattern speed $\Omega_{\mathrm{b}}=1.3\times\Omega_0$ to $\Omega_{\mathrm{b}}=1.425\times\Omega_0$ from left to right for a bar with a negative $m=4$ component with amplitude $\alpha_{m=4}=-0.0002$ to $-0.0012$ top to bottom.}
\label{ampsneg}
\end{figure*}

Figure \ref{m4slow} shows the $UV$ plane in the Solar neighbourhood for a 5 kpc bar with pattern speed $\Omega_{\mathrm{b}}=1.3\times\Omega_0$ to $\Omega_{\mathrm{b}}=1.425\times\Omega_0$ from left to right for a pure $m=2$ bar model (upper row), a negative amplitude $m=4$ model (middle row) and a positive amplitude $m=4$ model (lower row), assuming a flat rotation curve and a velocity dispersion of $0.15\times v_0$. We set $\alpha_{m=4}=\pm0.0005$, which is lower than the measured values in Section \ref{simulation}. However, the simulations also have higher values for $\alpha_{m=2}$ and the ratio between $\alpha_{m=2}$ and $\alpha_{m=4}$ (around $5-10$\%) is consistent. The upper row of Figure \ref{m4slow} shows that a long slow bar of 5 kpc with only an $m=2$ component does not reproduce a Hercules like feature within the range $\Omega_{\mathrm{b}}=1.3-1.425\times\Omega_0$. There is a slight deformation of the contours in the region of Hercules for $\Omega_{\mathrm{b}}=1.3-1.35\times\Omega_0$ arising from resonant trapping by the CR \citep{P-VPWG17}, but no clearly separated modes which would reproduce the stream. At higher pattern speeds a feature develops at $V\sim30$ km s$^{-1}$ arising from resonant trapping by the OLR, which is not seen in the Solar neighbourhood. Both these features are known and explained in more detail in \cite{2017MNRAS.465.1443M}. 

The second row of Figure \ref{m4slow} shows the $UV$ plane in the Solar neighbourhood for a 5 kpc bar comprised of a $m=2$ component with $\alpha_{m=2}=0.01$ and a negative $m=4$ component with $\alpha_{m=4}=-0.0005$. The negative $m=4$ component clearly creates a new divide in the $UV$ plane, which at higher pattern speeds is close to the observed location of Hercules. However, the bimodality only occurs in the right region to reproduce Hercules at pattern speeds where the OLR feature is also present at high $V$. 

The lower row of Figure \ref{m4slow} shows the $UV$ plane in the Solar neighbourhood for a 5 kpc bar comprised of a $m=2$ component with $\alpha_{m=2}=0.01$ and a positive $m=4$ component with $\alpha_{m=4}=0.0005$. The positive $m=4$ component also creates a new divide in the $UV$ plane, but it produces a feature at positive $U$, which is incompatible with the Hercules stream.

\begin{figure}
\centering
\includegraphics[width=\hsize]{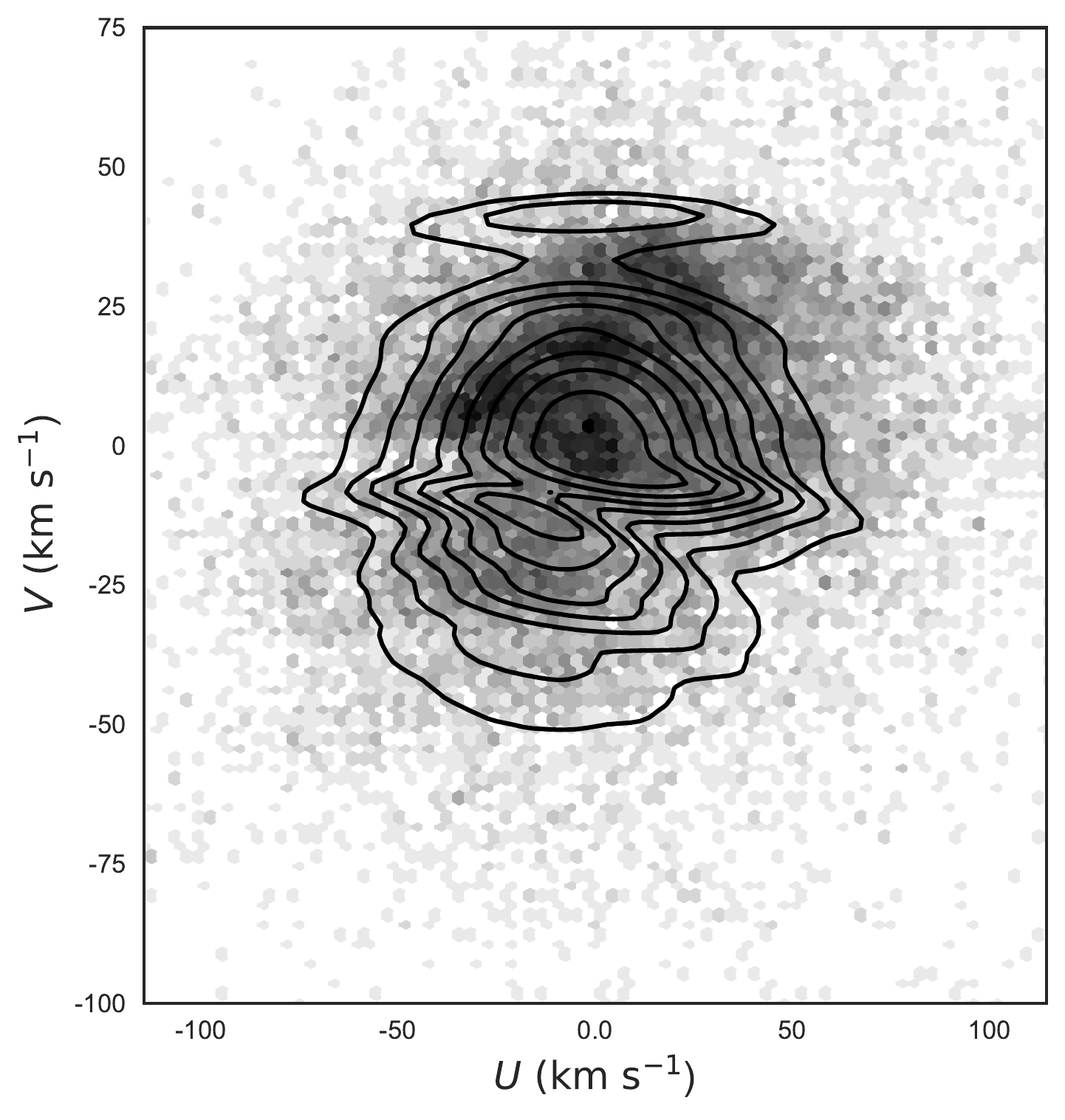}
\caption{$UV$ plane for the long slow bar model with $\Omega_{\mathrm{b}}=1.425\times\Omega_0$, $\alpha_{m=2}=0.01$ and $\alpha_{m=4}=-0.0005$ (contours) overlaid on the TGAS-RAVE $UV$ plane (hexbin).}
\label{425tgas}
\end{figure}

The amplitude of the $m=4$ component also has a significant effect on the $UV$ plane. Figure \ref{ampsneg} shows the $UV$ plane in the Solar neighborhood for a 5 kpc bar with pattern speed $\Omega_{\mathrm{b}}=1.3\Omega_0$ to $\Omega_{\mathrm{b}}=1.425\Omega_0$ from left to right for a bar with $\alpha_{m=2}=0.01$ and $\alpha_{m=4}=-0.0002$ to $-0.0012$ (top to bottom). When comparing the columns the progression of the resonance feature to lower $V$ with a higher pattern speed is clear, and to be expected. When comparing the rows the progressively stronger $m=4$ component leads to a larger deviation from the rough symmetry around $U=0$ of a pure $m=2$ bar. Once we approach $\alpha_{m=4}\approx-0.0008$ the perturbation in the contours of the velocity distribution becomes significantly less smooth, especially for the higher pattern speeds.

We find an amplitude of around $\alpha_{m=4}=-0.0005$ to be the best choice to reproduce a Hercules like feature, without causing a large disruption to the velocity distribution. Figure \ref{425tgas} shows contours from the long slow bar model with $\Omega_{\mathrm{b}}=1.4\times\Omega_0$, $\alpha_{m=2}=0.01$, $\alpha_{m=4}=-0.0005$ and a velocity dispersion of $0.13\times v_0$ overlaid on the TGAS-RAVE $UV$ plane correcting for the Solar motion using the values $U_{\odot}=-10$ km s$^{-1}$ \citep{Betal12} and $V_{\odot}=24$ km s$^{-1}$ \citep{2015ApJ...800...83B}. For the chosen values of the Solar motion the resonance of the $m=4$ component produces a Hercules like feature around $(U,V)=(-20,-20)$ km s$^{-1}$. However, the high $V$ feature arising from the OLR is not observed in the Solar neighbourhood $UV$ plane. We can reduce the intensity of the OLR feature by reducing the strength of the $m=2$ component of the bar. However, we show here the model with $\alpha_{m=2}=0.01$ which is a standard choice for the Dehnen bar model \citep[e.g.][]{D00}.  

This OLR feature around $V\sim30$ km s$^{-1}$ is an issue for a long bar model with $\Omega_{\mathrm{b}}\gtrsim1.35\times\Omega_0$. Unless some other component of the potential can be shown to suppress this response (e.g. interaction between the bar and spiral resonances) the lack of any observation of a similar feature in the Solar neighbourhood argues against a 5 kpc bar model with a pattern speed much above $\Omega_{\mathrm{b}}=1.35\times\Omega_0$. 

It is possible that a small OLR feature from a long slow bar at high $V$ has been observed in the TGAS proper motion data \citep{HKMGFS17} although it was interpreted as a signature of the Perseus arm in that work. It would require data at higher distances to trace whether the feature increases in strength with distance. Similarly, a small high $V$ feature is visible in the APOGEE2-S data \citep[visible in Fig. 4 of][]{Hunt+18} along the line of sight $(l,b)=(270^{\circ},0)$, although both features are low significance. If this is what has been observed, it would fit with a weaker bar, or a pattern speed around $\Omega_{\mathrm{b}}=1.35\times\Omega_0$.

\section{Four fold symmetry}\label{4sym}
If the Hercules stream originates from the resonance of a bar with an $m=4$ component then similar kinematics should be observable in four locations around the disc. E.g., assuming the Sun lies at $\phi=0^{\circ}$, then the $UV$ plane at $\phi=90^{\circ}$, $180^{\circ}$ and $270^{\circ}$ should contain similar resonance features from the $m=4$ component (it will only be identical at $\phi=180^{\circ}$ owing to the primary $m=2$ mode of the bar). In contrast, if the Hercules stream originates from the $m=2$ short fast bar model or the $m=2$ component of a long slow bar, then similar kinematics should only occur at $\phi=180^{\circ}$. 

Thus, if we can make observations across the disc we can potentially determine between Hercules resulting from a $m=2$ or $m=4$ component pattern. In turn, if we can make that distinction, we can also put a strong constraint on the pattern speed of the bar.

\cite{Bovy10} made a prediction across the disc for the Hercules stream if it originates from the short fast bar model. Here we make the same prediction around the $R_0=8$ kpc circle for the long slow bar with only an $m=2$ component with $\Omega_{\mathrm{b}}=1.3\times\Omega_0$, and one which includes an $m=4$ component with $\Omega_{\mathrm{b}}=1.4\times\Omega_0$ as shown in Figure \ref{425tgas}.

\begin{figure*}
\centering
\includegraphics[width=\hsize]{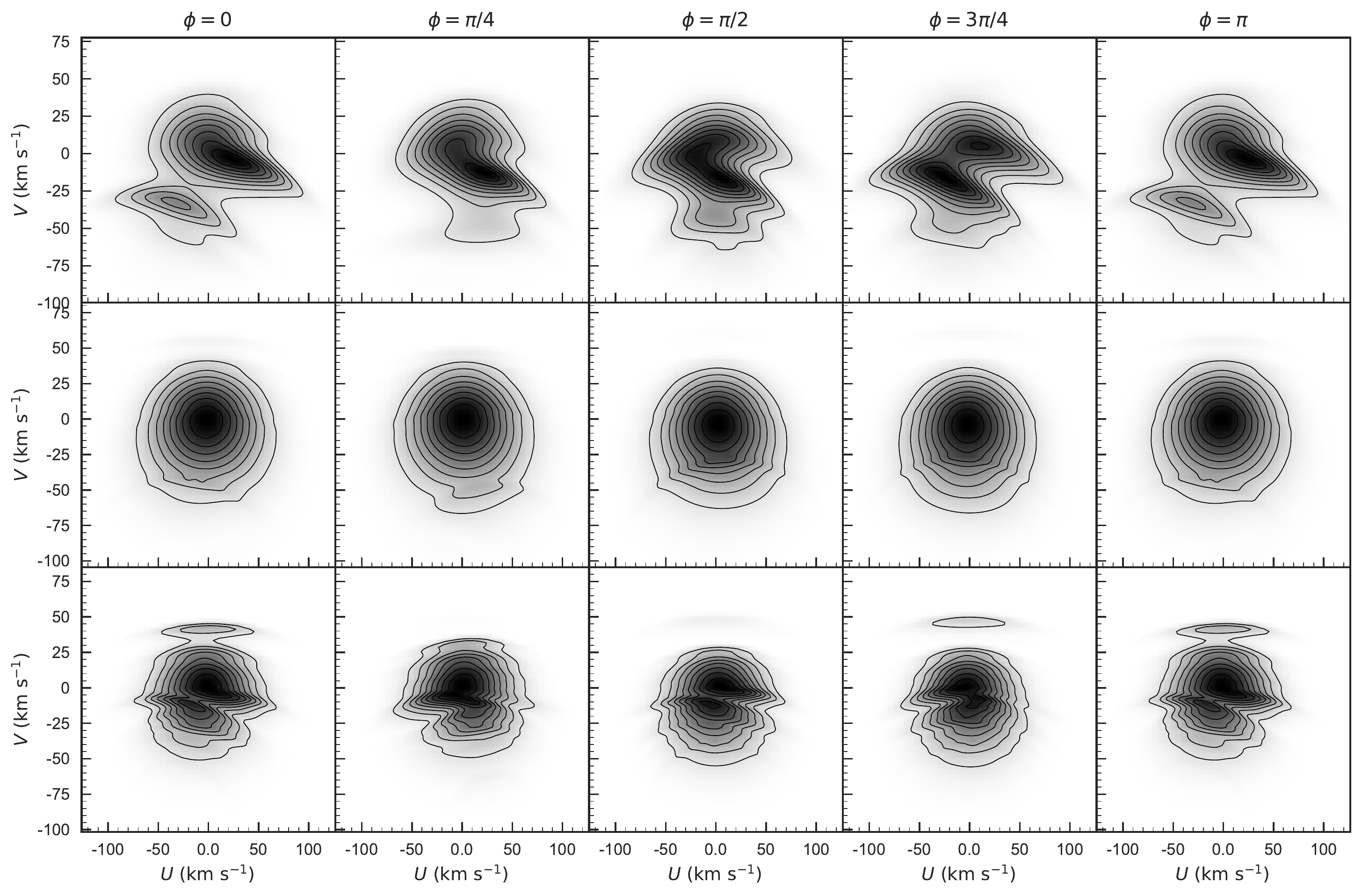}
\caption{$UV$ plane along the Solar circle ($R=8$ kpc) at varying azimuths by column for the short bar (top row), $m=2$ component long slow bar (centre row) and the $m=4$ long slow bar model from Figure \ref{425tgas} (bottom row).}
\label{symmetry}
\end{figure*}

Figure \ref{symmetry} shows the $UV$ plane along the Solar circle ($R=8$ kpc) at varying azimuths by column for the short bar model from \cite{Hunt+18} (top row), the $m=2$ component long slow bar model (centre row) where it is proposed the Hercules can originate from trapping around the CR, and the $m=4$ long bar model from Figure \ref{425tgas} (bottom row) where it is proposed that Hercules can originate from the 4:1 OLR. 

The top row of Figure \ref{symmetry} shows the $UV$ plane for an example short bar model. It displays two fold symmetry, e.g. the $UV$ plane at 0 and $\pi$ rad is identical. The second row of Figure \ref{symmetry} shows that the $UV$ plane of a standard Dehnen bar with $\alpha_{m=2}=0.01$ and $\Omega_{\mathrm{b}}=1.3\times\Omega_0$ also displays two fold symmetry. The third row of Figure \ref{symmetry} shows the $UV$ plane for a 5 kpc bar with $\alpha_{m=2}=0.01$, $\alpha_{m=4}=-0.0005$ and $\Omega_{\mathrm{b}}=1.425\times\Omega_0$. This bar models shows a mix of two and four fold symmetry. The OLR feature at high $V$ which originates from the $m=2$ component is two fold symmetric, such that it repeats every $\pi$ radians. However, the feature around the area of Hercules arising from the $m=4$ component is four fold symmetric, such that it repeats every $\pi/2$ radians, visible at $\phi=0, \pi/2$ and $\pi$ rad. 

In the near future data release 2 \citep[DR2;][]{KB17} from the European Space Agency's $Gaia$ mission \citep{GaiaMission} will provide detailed positions and proper motions for over $1.3\times10^9$ stars, and radial velocities for over $6\times10^6$ stars. This will allow us to explore kinematics away from the Solar neighbourhood and search for two or four fold symmetric features. We will not be able to examine the $UV$ plane across the disc specifically with DR2 owing to the lower number of radial velocities, but similar to the line-of-sight identified in \cite{Bovy10} which showed a strong signature of Hercules, there will exist lines of sight for which the gaps in the velocity distribution owing to various resonances are visible in the proper motion data alone. 

For that reason we do not make a detailed prediction of the model parameters, but merely highlight another potential mechanism for the creation of the Hercules stream in a model with a long-slow bar which can be tested against the data from $Gaia$ and other Galactic surveys. 

\section{Discussion and outlook}\label{summary}
In this work we have shown that it is possible to create a Hercules like feature in the Solar neighbourhood $UV$ plane in a model with a 5 kpc bar, containing both a $m=2$ and $m=4$ Fourier component. The other moving groups present in the main mode of the velocity distribution are not reproduced, likely because they originate from spiral resonances \citep[e.g.][]{QM05,Sellwood2010}, or interaction between the bar and spiral resonances \citep[e.g.][]{Quillen03,MFSGKB16}. We will investigate this further in an upcoming work (Hunt et. al, In prep).

Although existing long bar models have been able to reproduce a Hercules like feature through resonant trapping around the corotation radius \citep{P-VPWG17,MFFB17}, the effect is weaker than observed in the data. We are not suggesting that the parameterization of this model is a perfect representation of the bar, considering the OLR feature present at high $V$, but merely that a resonance origin for Hercules is compatible with the 4:1 OLR of a long bar model, providing a more complex potential is used. 

We make a general prediction of four fold symmetry across the Galactic disc if Hercules is caused by the $m=4$ component of the bar. At this stage we lack sufficient data to fully trace the stream to larger distances. However, in the near future $Gaia$ DR2 will provide detailed 5D phase space information for over $1.5\times10^9$ stars down to $\sim20$ mag, allowing us to trace how Hercules varies across the Galactic disc. This will enable us to make detailed comparisons with the competing models, and help explain the origins of the kinematic features in the Solar neighbourhood.

\section*{Acknowledgements} JH is supported by a Dunlap Fellowship at the Dunlap Institute for Astronomy \& Astrophysics, funded through an endowment established by the Dunlap family and the University of Toronto. JB received partial support from the Natural Sciences and Engineering Research Council of Canada. JB also received partial support from an Alfred P. Sloan Fellowship. This work has made use of data from the European Space Agency (ESA) mission Gaia (https://www.cosmos.esa.int/gaia), processed by the Gaia Data Processing and Analysis Consortium (DPAC, https://www.cosmos.esa.int/web/gaia/dpac/consortium). Funding for the DPAC has been provided by national institutions, in particular the institutions participating in the Gaia Multilateral Agreement. This work (or a part of this work) used the UCL facility Grace and the DiRAC Data Analytic system at the University of Cambridge, operated by the University of Cambridge High Performance Computing Service on behalf of the STFC DiRAC HPC Facility (www.dirac.ac.uk). This equipment was funded by BIS National E-infrastructure capital grant (ST/K001590/1), STFC capital grants ST/H008861/1 and ST/H00887X/1, and STFC DiRAC Operations grant ST/K00333X/1. DiRAC is part of the National E-Infrastructure.

\bibliographystyle{mn2e}
\bibliography{ref2}

\end{document}